\documentclass[reprint,aps]{revtex4-1}
\usepackage{amsmath}

\usepackage{graphicx}
\usepackage{dcolumn}
\usepackage{bm}
\usepackage{natbib}
\usepackage{pgfplots}

\begin{document}

\title{Velocity Statistics of the Nagel-Schreckenberg Model}

\author{Nicolas Bain}
\email{nbain@mit.edu}
\affiliation{Multi-Scale Materials Science for Energy and Environment, $\langle$MSE$\rangle^2$, UMI 3466, The joint CNRS-MIT Laboratory,
 Massachusetts Institute of Technology,
77 Massachusetts Avenue,
Cambridge, 02139, MA, USA}

\author{Thorsten Emig}
\email{emig@mit.edu}
\affiliation{Multi-Scale Materials Science for Energy and Environment, $\langle$MSE$\rangle^2$, UMI 3466, The joint CNRS-MIT Laboratory,
 Massachusetts Institute of Technology,
77 Massachusetts Avenue,
Cambridge, 02139, MA, USA}

\author{Michael Schreckenberg}
\email{michael.schreckenberg@uni-due.de}
\affiliation{Physik von Transport und Verkehr, Universit\"at Duisburg-Essen,
47048 Duisburg, Germany}

\author{Franz-Joseph Ulm}
\email{ulm@mit.edu}
\affiliation{Multi-Scale Materials Science for Energy and Environment, $\langle$MSE$\rangle^2$, UMI 3466, The joint CNRS-MIT Laboratory,
 Massachusetts Institute of Technology,
77 Massachusetts Avenue,
Cambridge, 02139, MA, USA}

\date{\today }

\begin{abstract}
  The statistics of velocities in the cellular automaton model of
  Nagel and Schreckenberg for traffic \cite{nagel1992cellular} are
  studied. From numerical simulations, we obtain the probability
  distribution function (PDF) for vehicle velocities and the
  velocity-velocity (vv) correlation function.  We identify the
  probability to find a standing vehicle as a potential order parameter that signals
  nicely the transition between free congested flow for sufficiently
  large number of velocity states. Our results for the vv correlation
  function resemble features of a second order phase transition. We
  develop a $3$-body approximation that allows us to relate the PDFs
  for velocities and headways. Using this relation, an approximation
  to the velocity PDF is obtained from the headway PDF observed in
  simulations. We find a remarkable agreement between this
  approximation and the velocity PDF obtained from simulations.
\end{abstract}

\pacs{Valid PACS appear here}
\maketitle

\section{\label{sec:Intro} Introduction}

Traffic flow theory is the backbone for understanding and improving the
mobility of people and goods in our road network. The classical tool
to characterizing mobility (in either field tests or from simulations)
is by plotting flow vs. density in form of the so-called fundamental
diagram (Fig.\ref{fundia}) for both individual road segments and road networks
\cite{schadschneider2010stochastic}; to separate free flow patterns
below a critical density from congested flow patterns above.  While
empirical approaches, based on equilibrium concepts, link vehicle
speed to density, eventually enriched by information related to human
behavior \cite{chowdhury2000statistical}, the fundamental structure of
the transition from free to congested flow remains a topical issue,
which may ultimately reconcile the considerable scatter in field
experiments with traffic flow theory and simulation. This provides
ample motivation for us to study velocity distribution functions of a
specific class of traffic models, cellular automata models (CA), that
are known to exhibit two different phases (free and congested flow)
and a transition between them
\cite{gerwinski1999analytic}. Specifically, we herein investigate the
velocity distribution functions for a simple version of these models,
namely the Nagel-Schreckenberg (NaSch) model for one lane traffic
\cite{nagel1992cellular}. The model is based on the discretization of
the road into cells of the size of a single vehicle, and the whole
system is described as the ensemble
$(v_{1},...,v_{N},d_{1},...,d_{N})$ of velocities $v_{j}$ and headways
$d_{j}$ (number of empty cells in front of a vehicle) of $N$ vehicles
\cite{nagel1992cellular,schreckenberg1995discrete}.  The time
evolution of the vehicles' positions $x_{j}$ and velocities $v_{j}$
follows four distinct update rules:

\begin{enumerate}
\item[(1)] Acceleration: $v_{j}=\min (v_{j}+1,v_{max})$

\item[(2)] Deceleration: $v_{j}=\min (d_{j},v_{j})$

\item[(3)] Random deceleration: $v_{j}=\max (v_{j}-1,0)$ with a probability $%
p$

\item[(4)] Movement: $x_{j}\rightarrow x_{j}+v_{j}$
\end{enumerate}

Herein, the velocity $v_{max}$, a model parameter, corresponds to the
maximum velocity that a vehicle can reach when there are no slower vehicles
ahead. The stochastic parameter $p$ represents the probability that a
vehicle randomly slows down, and aims at capturing the lack of perfection in
human behavior. In practice, the parameters $v_{max}$ and $p$ are kept
constant, while the total density of vehicles $\rho $ is varied, which is
the number of vehicles divided by the number of cells. For convenience, the
physical dimensions of headway and vehicle position are expressed in unit of
cells, and the velocity in unit of cells per iteration time step; thus
omitting the time dimension so that velocities and distances have formally
the same dimension.

The NaSch model has been studied through mean-field (MF) theories for which
velocity distribution functions were computed~\cite%
{schreckenberg1995discrete}. Since MF theories fail to give simple and
accurate results for values of $v_{max}>2$, our study of the probability
distribution function (PDF) for velocities of the NaSch model is based on
both numerical simulations and exact results for a simplified 3-body
approximation. To obtain the PDF, we analyze in detail the $v_{max}+1$
different accessible single vehicle velocity states between $0$ and $v_{max}$%
. We choose a value of $v_{max}=10$ which ensures obtaining a sufficiently
large number of states; and which is of the order of magnitude of typical
highway speeds. The stochastic parameter is chosen to be $p=0.5$ and remains
constant throughout the study. In order to avoid finite size effects we run
simulations with periodic boundary conditions over $2\times 10^{4}$ cells
for $10^{6}$ iteration time steps to reduce the influence of transient
behavior. To probe the effect of different initial conditions we initialized the system in three distinct ways:

\begin{itemize}
\item[(I1)] Megajam: block of $N$ standing vehicles \cite%
{gerwinski1999analytic}

\item[(I2)] Equally spaced, standing vehicles \cite{barlovic1998metastable}

\item[(I3)] Equally spaced vehicles moving with $v_{max}$
\end{itemize}

Our results turned out to be independent of these initial conditions, unless
mentioned explicitly later in the text.

As already noted, the NaSch model exhibits two different phases (free and
congested flow) and a transition between them with a yet to be defined order
parameter \cite{schadschneider2010stochastic}. Specifically, in the
deterministic case ($p=0$), a sharp phase transition occurs at a critical
density $\rho _{c}=1/(v_{max}+1)$ that coincides with the density of maximum
flow $j=\rho \langle v\rangle $, where $\langle v\rangle $ is the mean
velocity of all vehicles, averaged over time. It has been a subject of
intense debate in recent literature if the NaSch model exhibits a similar
sharp transition in the presence of randomness ($p>0$). In Section \ref%
{sec:Part2} we shed new light on this still open key question by analyzing
the statistics of velocities. We identify an approximate ``order parameter'', and
study the \textquotedblleft transition\textquotedblright\ by looking at the
velocity-velocity correlation function. In Section \ref{sec:Part3} we show
that the PDF for velocities and headways cannot be obtained
straightforwardly from the parameters of the model since there are no
separate PDFs in the free and jammed flow. Based on this insight thus
gained, we develop, in Section \ref{sec:Part4}, an exact solution for a
3-body approximation. This approximation is employed in Section \ref%
{sec:Part5} to provide estimates for the velocity PDF from the knowledge of
the PDF for headways. The paper concludes with a summary and discussion of
our findings.

\section{\label{sec:Part2} Potential ``order parameter'' and velocity
correlations}

\begin{figure}[tbp]
\includegraphics[width = \columnwidth]{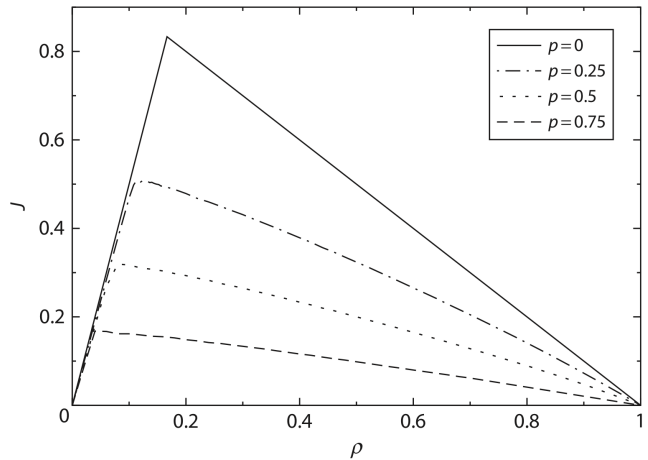} 
\caption{Numerically obtained fundamental diagrams of the NaSch model
  with $v_{max}=5$ for different values of the stochastic parameter
  $p$. (flow $j$ in arbitrary units.) 
Taken from Ref.~\cite{schadschneider2010stochastic}.}
\label{fundia}
\end{figure}

\begin{figure}[tbp]
\includegraphics[width = \columnwidth]{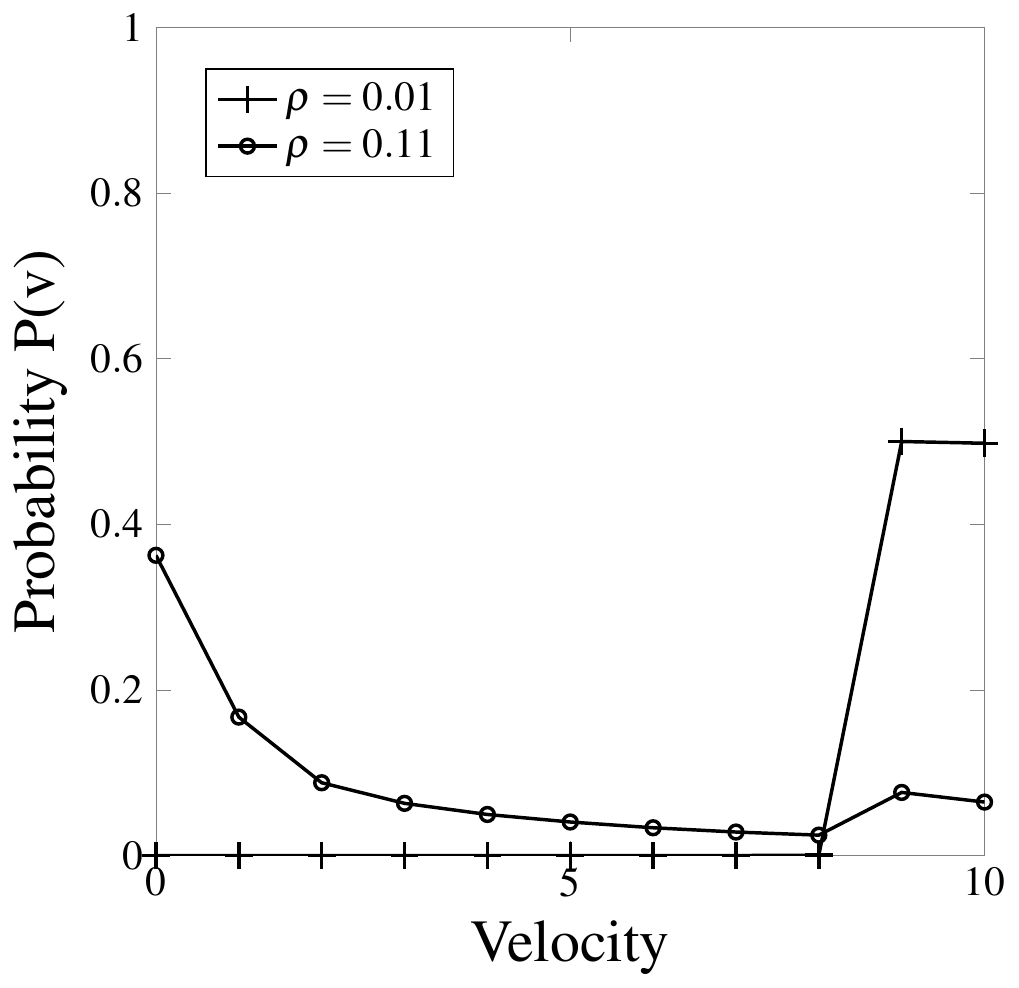} 
\caption{Velocity PDFs in both the free and congested flow phase. The PDF
for freely flowing vehicles is non-zero only for $v_{max}$ with value $1-p$
and $v_{max}-1$ with value $p$. In the congested regime the probability to
find a standing vehicle is finite, signaling the presence of traffic jams.}
\label{velDistrPlot}
\end{figure}

\begin{figure}[tbp]
\includegraphics[width = \columnwidth]{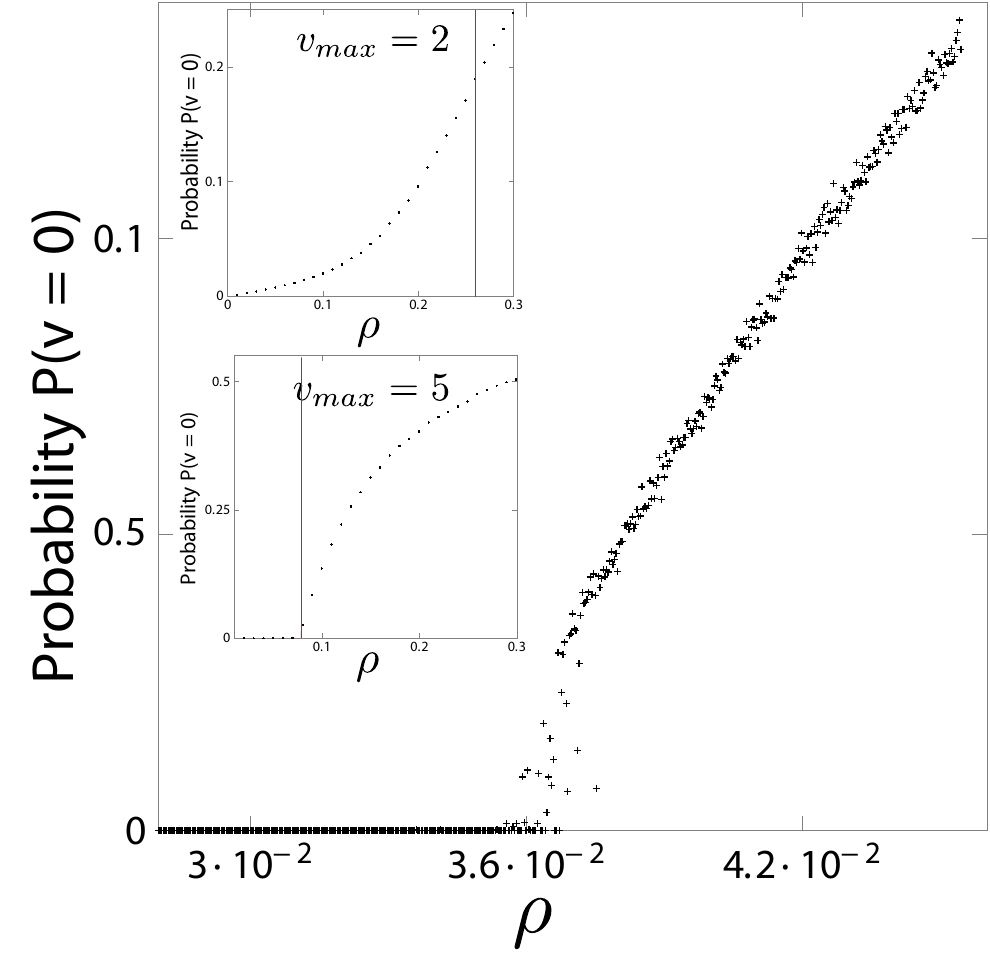} 
\caption{Probability to find a standing vehicle, $P(v=0)$ as function
  of the average density, for a system initialized with equally
  spaced, standing vehicles, and $v_{max}=10$. The critical density is
  close to $\protect\rho_c = 0.036$. Insets: $P(v=0)$ for two smaller
  $v_{max}$ with the critical densities obtained numerically in
  Ref.~\cite{gerwinski1999analytic} marked by vertical lines.}
\label{orderParameter}
\end{figure}

The NaSch model with randomness exhibits some kind of transition when
going from low to high densities. Below the transition, there exists a
free flow regime in which the interactions between vehicles are
negligible and the flow $j$ increases linearly with density. Above the
transition, one encounters a congested regime in which the flow $j$
decreases with increasing density. The flow-density relation, usually
referred to as fundamental diagram, hence clearly shows a
transition. Examples for the NaSch model for $v_{max}=5$ and different
values of the stochastic parameter $p$ are shown in Fig.~\ref{fundia}.
One might expect the existence of a genuine phase transition but it is
not straightforward to define a corresponding order parameter because
it is not obvious what symmetry is broken at the transition. To
address this point, we inspect the PDF for velocities, $P(v)$.
Averaging over all simulation time steps, we find two very distinct
distributions in the two phases (Fig.~\ref{velDistrPlot}). In the free
flow phase, the vehicles do not interact and every vehicle can travel
either at velocity $v_{max}$ or $v_{max}-1$ with some negligible
interactions. This solution is exactly known from the update rules,
from taking the limit $%
d_{j}>v_{max}$.
In the congested phase the probability to find a standing vehicle is
strictly non-zero, $P(v=0)>0$, and the probability to have a vehicle
at velocity $v_{max}$ or $v_{max}-1$ decays with increasing density.
The numerical result for $P(v=0)$ as a function of density is shown in
Fig.~\ref{orderParameter} for the initial condition (I2). The curve
shows a clear drop to zero at a density $\rho _{c}\approx 0.036$.
This suggest that $P(v=0)$ could be an order parameter for a putative
phase transition between free and congested flow. However, it has been
argued in Ref.~\cite{gerwinski1999analytic} that for small densities
$\rho\to 0$ one has the scaling $P(v=0) \sim \rho^{v_{max}-1}$ showing
that $P(v=0)$ is non-zero at {\it any} density, and hence cannot serve
as an order parameter in general. This can be reconciled with our
observation for $v_{max}=10$ by studying the behavior of $P(v=0)$ for
smaller $v_{max}$ which is shown in the insets of
Fig.~\ref{orderParameter}.  While for $v_{max}=2$ there is indeed a
continuos variation of $P(v=0)$ with a change of curvature close to
the critical density (density of maximum flow) found in
Ref.~\cite{gerwinski1999analytic}, already for $v_{max}=5$ there is a
rather steep drop in $P(v=0)$ very close to the critical density of
Ref.~\cite{gerwinski1999analytic}, see also Fig.~\ref{fundia} for
$p=0.5$. Our observations suggest that $P(v=0)$ can be considered as
an approximate or effective ``order parameter'' that describes the
transition rather well for sufficiently large $v_{max}$. We argue that
in a continuum version of the NaSch model with an infinite number
$v_{max}\to\infty$ of discrete velocity states, $P(v=0)$ becomes a
genuine order parameter.  This order parameter is different from
previously adopted choices such as the number of vehicles in the high
local density phases~\cite{lubeck1998density}, the density of
nearest-neighbor pairs at rest \cite{chowdhury2000statistical}, or the
deviation of the mean velocity from the velocity of free-moving
vehicles \cite{gerwinski1999analytic}.  The data for $P(v=0)$ are the
only ones in our numerical study that slightly depend on the initial
conditions, presumably due to the vicinity to the critical
density. However, the results for $v_{max}=10$ always show a
discontinuity around $\rho _{c}$. The behavior could be an artefact of
finite simulation time, but this problem seems unavoidable because of
the divergence of the time needed to reach a steady state close to
$\rho _{c}$ \cite{gerwinski1999analytic}.

\begin{figure}[tbp]
\includegraphics[width = \columnwidth]{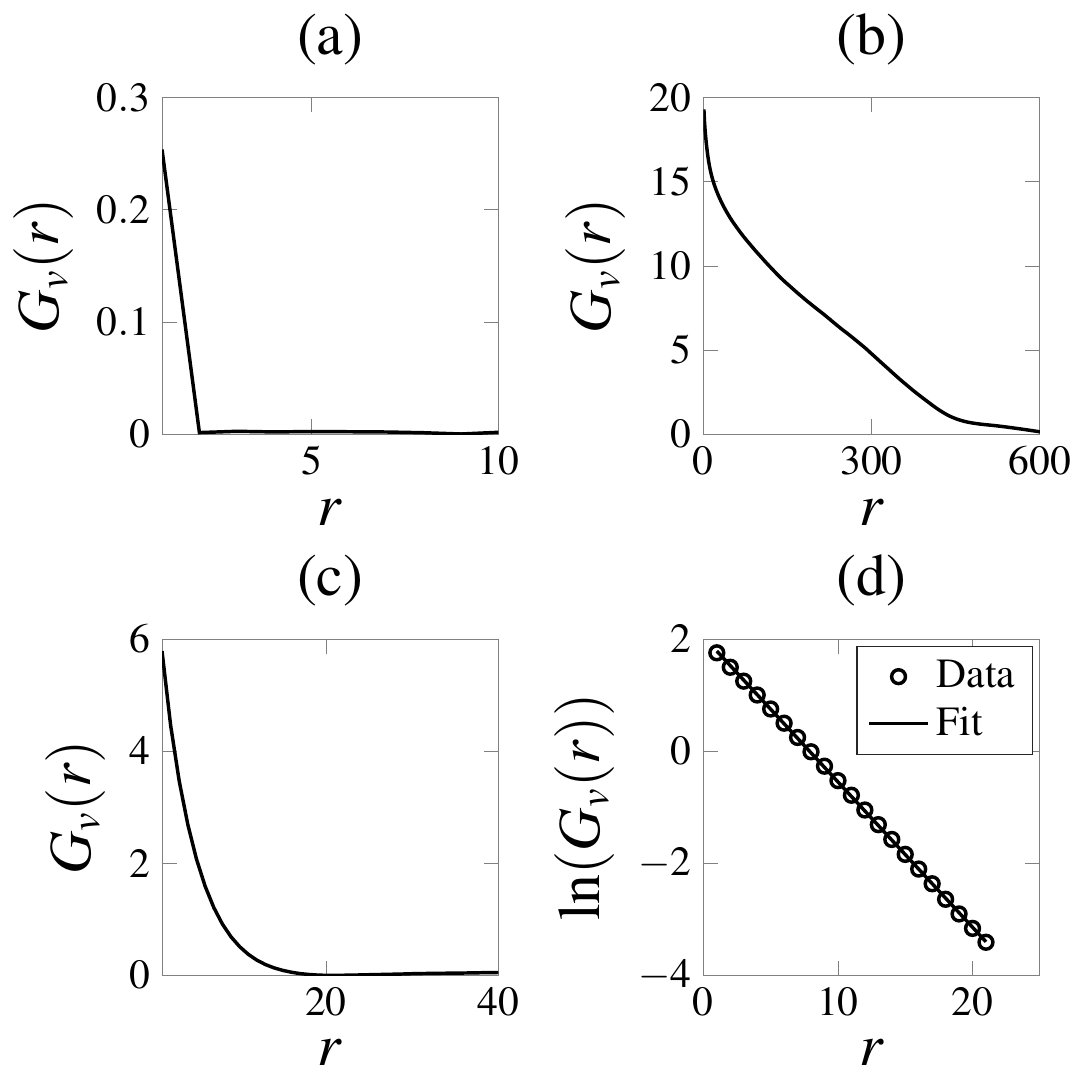} 
\caption{Velocity-velocity correlation function [see Eq.~(\protect\ref%
{velVelCor})] as a function of the vehicle number $r$ in the free flow
regime (a), just above the critical density $\protect\rho_c$ (b), and well
above the critical density (c). Shown in (d) is also a logarithmic plot of
the velocity-velocity correlation function well above the critical density
fitted with a line corresponding to a correlation number $r_c \approx 4$.}
\label{velCorrelationFunction}
\end{figure}

To gain a better understanding of the system's behavior at the
transition, we study the velocity-velocity correlation function
\begin{equation}
G_{v}(r)=\frac{1}{NT}\sum\limits_{t=1}^{T}\sum%
\limits_{j=1}^{N}v_{j}(t)v_{j+r}(t)-\langle v\rangle ^{2}  \label{velVelCor}
\end{equation}%
where $T$ is the simulation time, whereas $\langle v\rangle $ denotes the
average, over time and vehicles, of the individual vehicle velocities $%
v_{j}(t)$ at time step $t$:%
\begin{equation}
\langle v\rangle =\frac{1}{NT}\sum\limits_{j=1}^{N}\sum%
\limits_{t=1}^{T}v_{j}(t)\,,
\end{equation}%
We observe, in Fig.~\ref{velCorrelationFunction}, three distinct behaviors
of the velocity-velocity correlation function: (1) Well below the critical
density, at $\rho =0.01$, we find no correlation in the velocity of
successive vehicles [Fig.~\ref{velCorrelationFunction}(a)]. This is
consistent with a free flow in which all vehicles behave independently from
each another. (2) Well above the critical point, at $\rho =0.21$, [Fig.~\ref%
{velCorrelationFunction}(c)], we observe an exponential decay with a
correlation number of $r_{c}\approx 4$ successive vehicles [Fig.~\ref%
{velCorrelationFunction}(d)] which is characteristic of short-range
correlations. (3) Just above the critical density, at $\rho =0.05$, we find
a nearly linear decay of the correlation function, suggesting a diverging
correlation number $r_{c}$ [Fig.~\ref{velCorrelationFunction}(b)]. This
divergence of the correlation number close to the critical density mimics a
second-order phase transition.

\section{\label{sec:Part3} PDF of velocities}

\begin{figure}[tbp]
\includegraphics[width = \columnwidth]{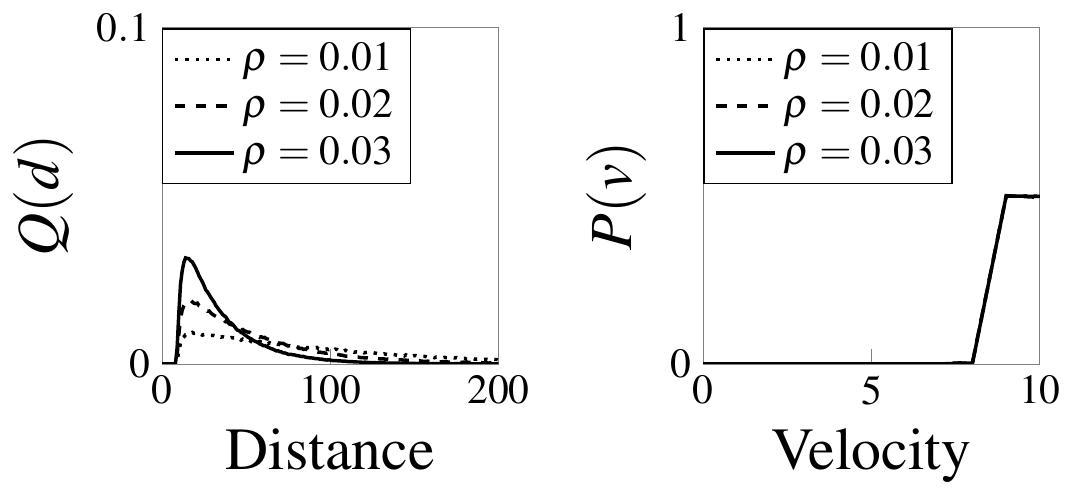} 
\caption{Velocity and headway PDFs for three different densities in the free
flow phase. The velocity distributions $P(v)$ are independent of these
densities but the headway distributions $Q(d)$ show a strong dependence on
density.}
\label{velAndNettoFF}
\end{figure}

The PDF of the velocities, as shown in Fig. \ref{velDistrPlot}, might
suggest that the PDFs for velocities and headways could be computed as the
sum of a PDF for the jammed phase ($P_{J}$ and $Q_{J}$, respectively) and a
PDF for free flow phase ($P_{F}$ and $Q_{F}$, respectively), weighted by the
proportion of the number of vehicles in each phase. This scheme has been
proposed in Ref.~\cite{krauss1996continuous}. When $\omega $ is the fraction
of vehicles in the free flow phase, this superposition assumption yields the
following PDF for velocities:  
\begin{equation}
P(v)=\omega P_{F}(v)+(1-\omega )P_{J}(v)\,.  \label{distrSeparation}
\end{equation}%
This form is convenient if one wants to predict velocity or headway PDFs
from a given density. To check the validity of this hypothesis, we computed
numerically the headway and velocity PDFs at different densities in two
regimes: below the critical density $\rho _{c}$ representative of the free
flow state; and well above the critical density $\rho _{c}$ where no
vehicles are in the free flow state. The resulting PDFs in the free flow
phase are shown in Fig.~\ref{velAndNettoFF}. According to Eq.~(\ref%
{distrSeparation}) the distributions should be independent of density. While
this seems to be true for the velocity PDF, for the distribution of headways
this is clearly not the case. This observation shows that the superposition
model expressed by (\ref{distrSeparation}) is but an oversimplification of the true
behavior of the model. It suggests that the internal structure of the
traffic state quite certainly plays a role in the shape of the headway and
velocity PDFs. One therefore needs a new method to predict velocity and
headway distributions for the NaSch model. 

\section{\label{sec:Part4} A $3$-body approximation: Exact solution}

We approach this problem analytically, by considering a $3$-body
approximation for the NaSch model for which an exact analytical solution is
derived. To this end, consider a standing vehicle which we number as $j=0$.
Assume that this vehicle represents the tail of a jam and that it remains
stopped. At a distance $d_{0}$ behind this vehicle we place a second
vehicle, $j=1$, and a third one, $j=2$ in the cell immediately behind
vehicle $j=1$, both initially standing. This is the initial configuration.
Then compute at every iteration step the joint PDF $%
P(d_{1},v_{1},d_{2},v_{2},t|d_{0})$ which is the probability to find at time 
$t$ vehicle $j=1$ at a distance $d_{1}$ from vehicle $j=0$ at velocity $v_{1}
$, and to find vehicle $j=2$ at a distance $d_{2}$ from vehicle $j=1$ and
velocity $v_{2}$. This joint PDF $P$ is calculated iteratively from the
initial conditions 
\begin{equation}
P(d_{1},v_{1},d_{2},v_{2},t|d_{0})=\delta _{d_{1},d_{0}}\delta
_{v_{1},0}\delta _{d_{2},0}\delta _{v_{2},0}
\end{equation}%
with $\delta _{a,b}$ the Kronecker delta. The time evolution of this PDF is
determined by iteration rules that depend on different regimes of distances
between the three vehicles. In the simplest case of sufficiently large
distances, with the definition $P_{a,b}=P(d_{1}+a,a,d_{2}-a+b,b,t-1|d_{0})$,
the iteration rule can be written as 
\begin{align}
P(d_{1},v_{1},d_{2},v_{2},t|d_{0})& =ppP_{v_{1},v_{2}}+pqP_{v_{1},v_{2}-1}
\label{iterationStep} \\
& +qpP_{v_{1}-1,v_{2}}+qqP_{v_{1}-1,v_{2}-1}  \nonumber
\end{align}%
with $q=1-p$. The precise form of the other iteration rules~(\ref%
{iterationStep}) slightly differ in the conditions on $d_{1}$, $v_{1}$, $%
d_{2}$ and $v_{2}$; and we refer for details to Appendix \ref{sec:annexA}.
From this joint distribution, the PDF of a given variable is obtained by
summing over all other variables. For instance, the velocity PDF $%
p_{1}(v_{1},t|d_{0})$ of vehicle $j=1$ at time $t$ reads as: 
\begin{equation}
p_{1}(v_{1},t|d_{0})=\sum\limits_{d_{1}=0}^{d_{0}}\sum%
\limits_{d_{2}=0}^{d_{0}}\sum%
\limits_{v_{2}=0}^{v_{max}}P(d_{1},v_{1},d_{2},v_{2},t|d_{0})\,.
\end{equation}%
The time evolution of this PDF for $d_{0}=500$ is plotted in Fig.~\ref%
{v1PdfTimeEvolution} for all $v_{max}+1$ velocity states. From the
probabilities for the different velocities $v_{1}$, we recognize three
distinct phases; namely acceleration, free flow, and deceleration. 
\begin{figure}[tbp]
\includegraphics[width = \columnwidth]{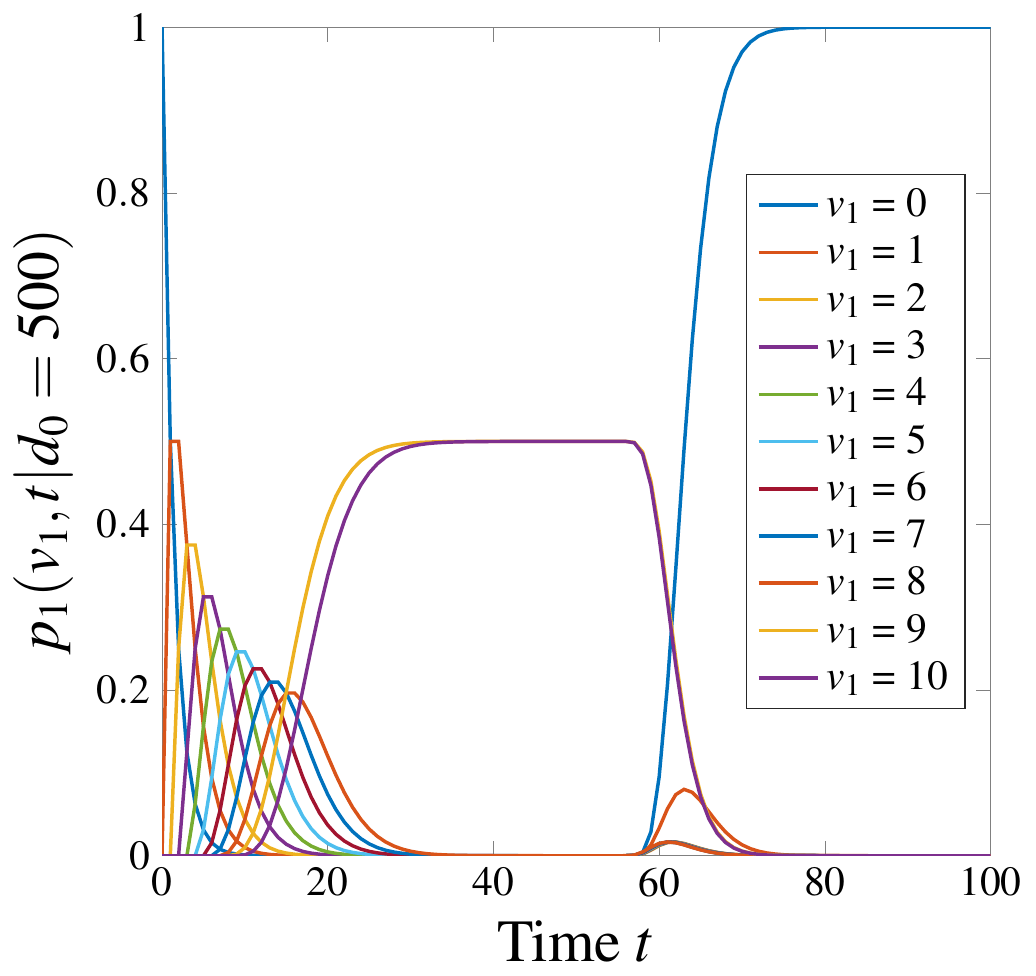} 
\caption{Time evolution of the PDF $p_{1}(v_{1},t|d_{0})$ of vehicle $1$ for
all possible velocities and for an initial spacing $d_{0}=500$. We notice
three distinct regimes in the time evolution of the probability to find
vehicle $1$ at velocity $v_{1}$. First the vehicle accelerates,
corresponding to a decrease of the probability that it stands. Next, the
occurence of finite probabilities for intermediate velocities, signalling a
transition to free flow with equally likely velocities $v_{max}-1$ and $%
v_{max}$. Eventually, deceleration when vehicle $1$ approaches the tail of
the jam, corresponding to probability $1$ to find the vehicle standing. }
\label{v1PdfTimeEvolution}
\end{figure}

\section{\label{sec:Part5}Application: relating headway and velocity
distributions}

\begin{figure}[tbp]
\includegraphics[width = \columnwidth]{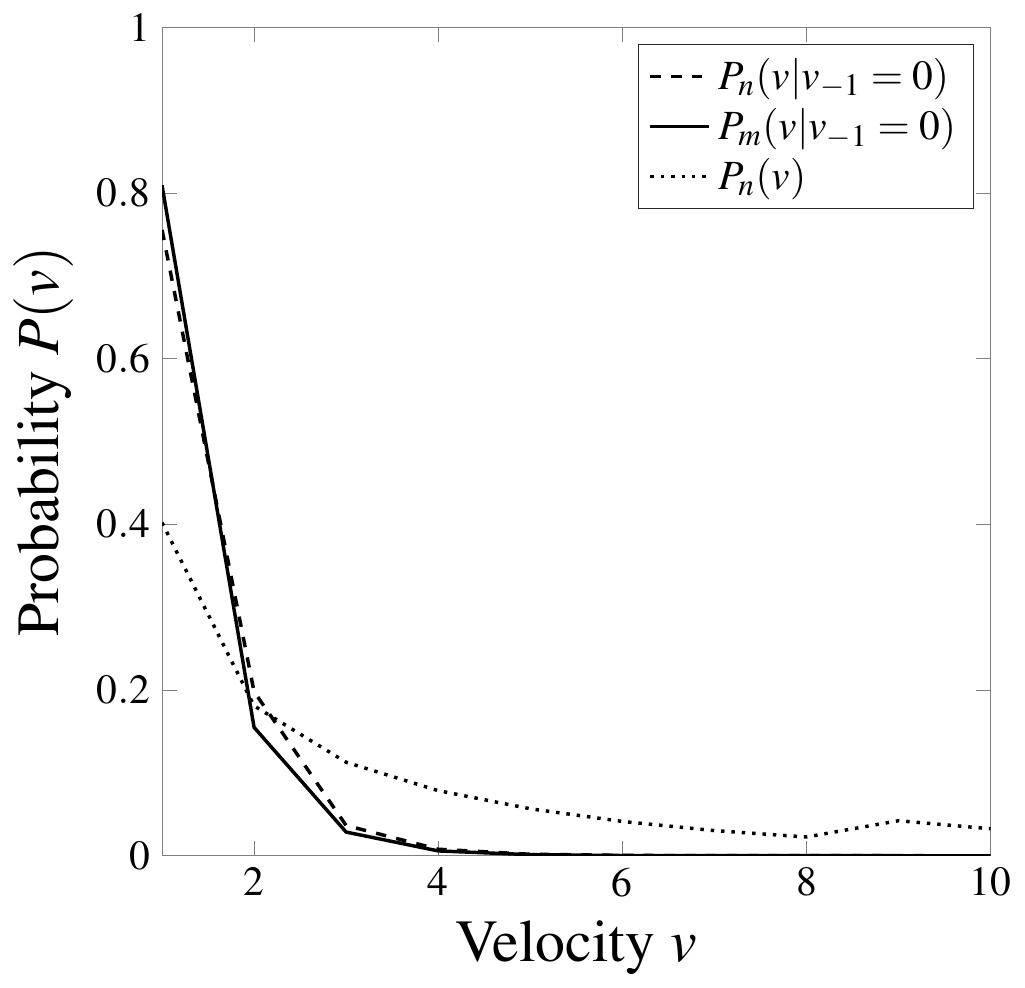} 
\caption{Velocity PDFs $P_n$ obtained numerically in the constrained and
unconstrained case, and the velocity PDF $P_m$ obtained from the headway
distribution and the three-body approximation in the constrained case. The
density is fixed at $\protect\rho = 0.21$.}
\label{velDistrInFrontOfStopped}
\end{figure}

In the previous section we developed an exact solution for PDFs within a $3$%
-body approximation. We now investigate how this result can be used to link
headway and velocity PDFs in the congested state. For this purpose, we base
the reasoning on our observation that in a congested state the probability
to find a standing vehicle is non-vanishing, see Section \ref{sec:Part2}.
Therefore, if one takes a snapshot of a congested state at a given time, one
observes standing vehicles with moving vehicles in between them. We start by
focusing on these standing vehicles and the ones directly ahead of them, and
hence on the constrained headway PDF of any stopped vehicle $%
Q_{n}(d_{i}|v_{i}=0)$, and the constrained velocity PDF of any vehicle which
is followed by a stopped vehicle $P_{n}(v_{j}|v_{j+1}=0)$. While these two
distributions can be obtained from our numerical simulation results with the
NaSch model, we herein aim at obtaining an approximation $%
P_{m}(v_{1}|v_{2}=0)$ for this constrained velocity PDF within the $3$-body
approximation. We thus relate $P_{m}(v_{1}|v_{2}=0)$ to the numerically
obtained headway PDF $Q_{n}(d_{i}|v_{i}=0)$. By comparing the approximate to
the numerically obtained velocity PDF, we can probe the accuracy of the $3$%
-body approximation. In order to relate the numerically computed headway
distribution $Q_{n}(d_{i}|v_{i}=0)$ to the $3$-body approximation, we assume
that a congested state can be described as a succession of $3$-body blocks
with a PDF $Q_{0}(d_{0})$ for the initial spacing $d_{0}$. We define the
summed probabilities 
\begin{equation}
\begin{split}
D_{m}(d_{2},d_{0})&
=\sum\limits_{t=0}^{T}\sum\limits_{d_{1}=0}^{d_{0}}\sum%
\limits_{v_{1}=0}^{v_{max}}P(d_{1},v_{1},d_{2},0,t|d_{0}) \\
V_{m}(v_{1},d_{0})&
=\sum\limits_{t=0}^{T}\sum\limits_{d_{1}=0}^{d_{0}}\sum%
\limits_{d_{2}=0}^{d_{0}}P(d_{1},v_{1},d_{2},0,t|d_{0})
\end{split}
\label{matrixForm}
\end{equation}%
which can be viewed as matrices with row index $d_2$ (or $v_1$) and
column index $d_{0}$.  This allows us to express the headway
distribution $Q_{m}(d_{2}|v_{2}=0)$ of vehicle $2$ when it remains
standing as
\begin{equation}
Q_{m}(d_{2}|v_{2}=0)=D_{m}.Q_{0}=Q_{n}(d_{i}|v_{i}=0)
\label{QEstimateConstrained}
\end{equation}%
Similarly, we can easily obtain the velocity PDF of vehicles followed by a
standing vehicle as 
\begin{equation}
P_{m}(v_{1}|v_{2}=0)=V_{m}.Q_{0}=V_{m}.(D_{m})^{-1}.Q_{n}(d_{i}|v_{i}=0)
\label{velEstimateConstrained}
\end{equation}%
Note that the matrix $D_{m}$ is invertible because it is triangular with
non-vanishing diagonal elements. A comparison of the constrained velocity
PDF obtained respectively from simulations, $P_{n}(v_{j}|v_{j+1}=0)$, and
analytically with the $3$-body approximation, $P_{m}(v_{1}|v_{2}=0)$, is
plotted in Fig.~\ref{velDistrInFrontOfStopped} for a density of $\rho =0.21$%
, and shows a fairly good agreement achieved without any adjustable fitting
parameters. Worthnoty in this plot is the significant difference between the
constrained and the unconstrained velocity PDFs, $P_{n}(v_{j}|v_{j+1}=0)$
and $P_{n}(v)$, which underlines the limitations of mean-field theories.
Note also, that in order to keep the total iteration time $T$ small
enough, we use the velocity and headway distributions for moving vehicles only,
i.e., excluding zero velocity and headway. 

\begin{figure}[tbp]
\includegraphics[width = \columnwidth]{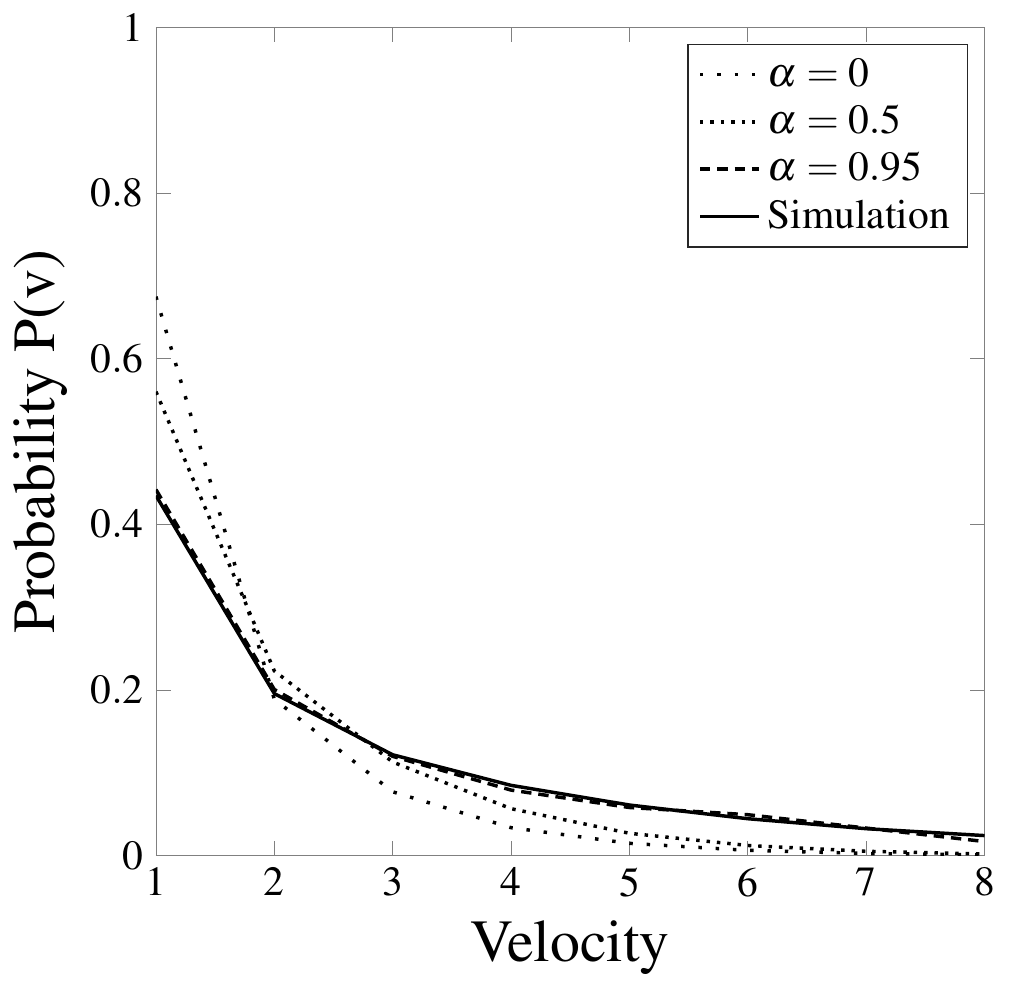} 
\caption{Approximate results for the velocity PDF for different values of
the fitting parameter $\protect\alpha$ (dashed and dotted lines), and the
velocity PDF obtained from simulations for a density $\protect\rho = 0.21$.}
\label{velDistrFromExact}
\end{figure}

Next, we investigate how to relate the \textit{unconstrained} velocity and
headway PDFs that we obtained numerically, by using the three-body
approximation. To do so, one has to estimate the global headway PDF from the
headway distributions of the two vehicles $j=1,\,2$ of the $3$-body blocks.
First, we define the headway PDFs as the matrices 
\begin{equation}
\begin{split}
\hat{D}_{m,1}(d,d_{0})&
=\sum\limits_{t,v_{1},d_{2},v_{2}}P(d,v_{1},d_{2},v_{2},t|d_{0}) \\
\hat{D}_{m,2}(d,d_{0})&
=\sum\limits_{t,d_{1},v_{1},v_{2}}P(d_{1},v_{1},d,v_{2},t|d_{0})\,.
\end{split}%
\end{equation}%
Next, we introduce a parameter $\alpha $ to approximate the global headway
distribution as a superposition of the ones for the two vehicles, 
\begin{equation}
\hat{D}_{m}=(1-\alpha )\hat{D}_{m,1}+\alpha \hat{D}_{m,2}\,.
\end{equation}%
Finally, we define the velocity PDFs 
\begin{equation}
\begin{split}
\hat{V}_{m,1}=(v,d_{0})&
=\sum\limits_{t,d_{1},d_{2},v_{2}}P(d_{1},v,d_{2},v_{2},t|d_{0}) \\
\hat{V}_{m,2}=(v,d_{0})&
=\sum\limits_{t,d_{1},d_{2},v_{1}}P(d_{1},v_{1},d_{2},v,t|d_{0})\,.
\end{split}%
\end{equation}%
Since we have $\hat V_{m,1} \approx \hat V_{m,2}$, we will simply
set $\hat V_m = \hat V_{m,1}$. We note
that this approximation is justified since the velocity of
vehicle $j=2$ is essentially the same as the one of vehicle $j=1$, but
shifted in time.  This does not hold for the headway distributions
since the initial spacing between vehicles $j=1$ and $j=0$ is
significantly different from the one between vehicles $j=2$ and $j=1$.
Now we can use Eqs.~(\ref{QEstimateConstrained}),
(\ref{velEstimateConstrained}) with $D_m$ replaced by $\hat D_m$ and
$V_m$ replaced by $\hat V_m$ to obtain an approximation for the
unconstrained velocity PDF $P(v)$ from the simulation result for the
headway PDF $Q_n(d|v=0)$. This approximation is now compared to the
velocity PDF obtained from the simulations where $\alpha$ is used as a
fitting parameter.  The results are shown in
Fig.~\ref{velDistrFromExact} for the choices $\alpha = 0$,
$\alpha = 0.5$ and $\alpha = 0.95$ for a fixed density $\rho = 0.21$.
One observes a remarkable agreement between the PDFs obtained respectively from our
numerical simulations and the $3$-body
approximation applied to the simulation result for the headway
distribution when the fitting parameter is $\alpha = 0.95$.  The
physical meaning of the fitting parameter, however, is not immediately
obvious.  A possible interpretation is the following.  The basic
assumption for the approximation presented here is that a congested
state can be considered at any time step as a distribution of standing
vehicles with moving vehicles in between them.  In our $3$-body
approximation there is always a pair of moving vehicles ($j=1,\, 2$)
that approaches the tail of a jammed region (vehicle $j=0$).  Hence,
one can view $\alpha$ as the relative importance of the distribution
of the distance between the two vehicles of a pair, and $1-\alpha$ as
the relative importance of the distribution of the distance between a
moving pair and the tail of the jam ahead. Our comparison in
Fig.~\ref{velDistrFromExact} then suggests that the velocity PDF in a
jammed state is much more sensitive to the constrained distribution of
the distances between two moving vehicles approaching a standing
vehicle than the distribution of actual distances of the pair to the
standing vehicle. This in turn suggests that the velocity PDF in a
congested state is mostly determined by the dynamics within regions of
still moving vehicles and not the interaction of these regions with
jammed regions.

\section{\label{sec:Conclusion}Conclusion}

In this paper we analyzed the statistics of velocities in the NaSch
model.  We studied the characteristics of the two phases and the phase
transition in between them. We find that in the free flow phase the
interactions between vehicles are negligible and that the velocity PDF
assumes a simple form. The congested phase is characterized by a
non-zero probability to find a standing vehicle. We suggest to use
this probability as an approximate ``order parameter'' that might
become a genuine order parameter in a continuum version of the model
where $v_{max}\to\infty$ velocity states exist.  The nature of the
phase transition could not be fully determined. There seems to be a
discontinuity at a critical density but this could be an artefact of a
too short simulation time. The velocity-velocity correlation function
shows three different regimes: Free flow with no correlation in
velocity, a critical regime with diverging correlation number and
highly congested flow with a short correlation number. This scenario
resembles a second order phase transition. 

Our simulations show that all PDFs are highly dependent on the density,
whether in free flow or in the completely congested regime, with the
exception of a simple PDF for the velocities in the low-density free flow.
We presented an analytical solution for a three-body approximation, that
allowed us to analytically link the headway and velocity distribution
functions. We first focused on a constrained case with stopped vehicles, and
then studied the unconstrained case. This approximation describes nicely the
constrained case. By introducing a fitting parameter in the constrained
case, we could also reach a remarkable agreement between simulation results
and our approximation. An attempt to interpret the physical meaning of the
fitting parameter has been made. The method presented in this paper to link
headway and velocity PDFs could prove useful to link other quantities of
interest and to learn more about the local dynamics in the jammed phase.

In addition to offering a new perspective on traffic models, the
present study of velocity statistics might help identifying the nature
of a potential phase transition in the stochastic NaSch model.
Moreover, in more applied terms, the knowledge of velocity
distributions enables one to compute expected excess fuel consumption
related to road properties as the ones studied in
Refs. \cite{deflection,louhghalam2014scaling,roughness} as a function
of traffic conditions, which is the topic of a forthcoming paper.
This relation opens new perspectives, alongside with the recent boom
of data collections \cite{fazeen2012safe}, which enables one to
revisit traffic models and adapt them to meet the requirements of
other research areas like, e.g., carbon management.

\appendix

\section{\label{sec:annexA}Details of the $3$-body approximation}

In order to obtain the complete iteration rules of the $3$-body interaction
approximation we introduce the notion of \textit{limiting speed} $\hat v_j =
\min(d_j, v_{max})$, which is the speed that the vehicle $j$ cannot exceed
at any given time. For each vehicle we have to distinguish three cases:

\begin{itemize}
\item[(i)] $v_j < \hat v_j -1$

\item[(ii)] $v_j = \hat v_j -1$

\item[(iii)] $v_j = \hat v_j$
\end{itemize}

with the assumption that $d_j >1$, the case $d_j=0$ being trivial. The
indices $1$ and $2$ being interchangeable, we denote six different cases:

\begin{enumerate}
\item[(1)] $v_1 < \hat v_1 - 1$ and $v_2 < \hat v_2 - 1$%

\item[(2)] $v_1 = \hat v_1 - 1$ and $v_2 < \hat v_2 - 1$%

\item[(3)] $v_1 = \hat v_1$ and $v_2 < \hat v_2 - 1$

\item[(4)] $v_1 = \hat v_1 - 1$ and $v_2 = \hat v_2 - 1$%

\item[(5)] $v_1 = \hat v_1$ and $v_2 = \hat v_2 - 1$

\item[(6)] $v_1 = \hat v_1$ and $v_2 = \hat v_2$
\end{enumerate}

Finally we introduce the scalar variables: 
\begin{equation}
r_j = \left\lbrace 
\begin{array}{ll}
p \mbox{ if } v_j < \hat v_j &  \\ 
q \mbox{ if } v_j = \hat v_j & 
\end{array}
\right. \, ; \, s_j = \left\lbrace 
\begin{array}{ll}
q \mbox{ if } v_j > 0 &  \\ 
0 \mbox{ if } v_j = 0 & 
\end{array}
\right.
\end{equation}
With that in hand we write the iteration rule for all the cases. Case
(1) is the most simple and given in Eq. (\ref{iterationStep}). Cases (2) and
(3) can be grouped together and follow the iteration rule: 
\begin{align}
& P = r_1 \sum\limits_{w_1 = v_1}^{v_{max}} \left( r_2 P_{w_1,v_2} + s_2
P_{w_1, v_2-1} \right)  \nonumber \\
&\quad +s_1r_2P_{v_1-1, v_2} + s_1s_2 P_{v_1-1,v_2-1}
\end{align}
with $P = P(d_1, v_1, d_2, v_2, t |d_0)$. As for cases (4), (5) and (6),
they also can be grouped in the same iteration rule, in the form: 
\begin{align}
& P = r_1 \sum\limits_{w_1 = v_1}^{v_{max}} \left( r_2 \sum\limits_{w_2 =
v_2}^{v_{max}} P_{w_1,w_2} + s_2 P_{w_1, v_2-1} \right)  \nonumber \\
&\quad +s_1r_2\sum\limits_{w_2 = v_2}^{v_{max}}P_{v_1-1, w_2} + s_1s_2
P_{v_1-1,v_2-1}
\end{align}
Implementing these iteration rules numerically is tedious but not
technically difficult, and enables one to retrieve the results described in
Section (\ref{sec:Part4}).

\renewcommand\refname{Bibliography}

\bibliographystyle{apsrev4-1}
\bibliography{NaSchVelDistr}

\begin{thebibliography}{12}%
\makeatletter
\providecommand \@ifxundefined [1]{%
 \@ifx{#1\undefined}
}%
\providecommand \@ifnum [1]{%
 \ifnum #1\expandafter \@firstoftwo
 \else \expandafter \@secondoftwo
 \fi
}%
\providecommand \@ifx [1]{%
 \ifx #1\expandafter \@firstoftwo
 \else \expandafter \@secondoftwo
 \fi
}%
\providecommand \natexlab [1]{#1}%
\providecommand \enquote  [1]{``#1''}%
\providecommand \bibnamefont  [1]{#1}%
\providecommand \bibfnamefont [1]{#1}%
\providecommand \citenamefont [1]{#1}%
\providecommand \href@noop [0]{\@secondoftwo}%
\providecommand \href [0]{\begingroup \@sanitize@url \@href}%
\providecommand \@href[1]{\@@startlink{#1}\@@href}%
\providecommand \@@href[1]{\endgroup#1\@@endlink}%
\providecommand \@sanitize@url [0]{\catcode `\\12\catcode `\$12\catcode
  `\&12\catcode `\#12\catcode `\^12\catcode `\_12\catcode `\%12\relax}%
\providecommand \@@startlink[1]{}%
\providecommand \@@endlink[0]{}%
\providecommand \url  [0]{\begingroup\@sanitize@url \@url }%
\providecommand \@url [1]{\endgroup\@href {#1}{\urlprefix }}%
\providecommand \urlprefix  [0]{URL }%
\providecommand \Eprint [0]{\href }%
\providecommand \doibase [0]{http://dx.doi.org/}%
\providecommand \selectlanguage [0]{\@gobble}%
\providecommand \bibinfo  [0]{\@secondoftwo}%
\providecommand \bibfield  [0]{\@secondoftwo}%
\providecommand \translation [1]{[#1]}%
\providecommand \BibitemOpen [0]{}%
\providecommand \bibitemStop [0]{}%
\providecommand \bibitemNoStop [0]{.\EOS\space}%
\providecommand \EOS [0]{\spacefactor3000\relax}%
\providecommand \BibitemShut  [1]{\csname bibitem#1\endcsname}%
\let\auto@bib@innerbib\@empty
\bibitem [{\citenamefont {Nagel}\ and\ \citenamefont
  {Schreckenberg}(1992)}]{nagel1992cellular}%
  \BibitemOpen
  \bibfield  {author} {\bibinfo {author} {\bibfnamefont {K.}~\bibnamefont
  {Nagel}}\ and\ \bibinfo {author} {\bibfnamefont {M.}~\bibnamefont
  {Schreckenberg}},\ }\href@noop {} {\bibfield  {journal} {\bibinfo  {journal}
  {Journal de physique I}\ }\textbf {\bibinfo {volume} {2}},\ \bibinfo {pages}
  {2221} (\bibinfo {year} {1992})}\BibitemShut {NoStop}%
\bibitem [{\citenamefont {Schadschneider}\ \emph {et~al.}()\citenamefont
  {Schadschneider}, \citenamefont {Chowdhury},\ and\ \citenamefont
  {Nishinari}}]{schadschneider2010stochastic}%
  \BibitemOpen
  \bibfield  {author} {\bibinfo {author} {\bibfnamefont {A.}~\bibnamefont
  {Schadschneider}}, \bibinfo {author} {\bibfnamefont {D.}~\bibnamefont
  {Chowdhury}}, \ and\ \bibinfo {author} {\bibfnamefont {K.}~\bibnamefont
  {Nishinari}},\ }\href@noop {} {\emph {\bibinfo {title} {Stochastic transport
  in complex systems: from molecules to vehicles}}}\ (\bibinfo  {publisher}
  {Elsevier})\BibitemShut {NoStop}%
\bibitem [{\citenamefont {Chowdhury}\ \emph {et~al.}(2000)\citenamefont
  {Chowdhury}, \citenamefont {Santen},\ and\ \citenamefont
  {Schadschneider}}]{chowdhury2000statistical}%
  \BibitemOpen
  \bibfield  {author} {\bibinfo {author} {\bibfnamefont {D.}~\bibnamefont
  {Chowdhury}}, \bibinfo {author} {\bibfnamefont {L.}~\bibnamefont {Santen}}, \
  and\ \bibinfo {author} {\bibfnamefont {A.}~\bibnamefont {Schadschneider}},\
  }\href@noop {} {\bibfield  {journal} {\bibinfo  {journal} {Physics Reports}\
  }\textbf {\bibinfo {volume} {329}},\ \bibinfo {pages} {199} (\bibinfo {year}
  {2000})}\BibitemShut {NoStop}%
\bibitem [{\citenamefont {Gerwinski}\ and\ \citenamefont
  {Krug}(1999)}]{gerwinski1999analytic}%
  \BibitemOpen
  \bibfield  {author} {\bibinfo {author} {\bibfnamefont {M.}~\bibnamefont
  {Gerwinski}}\ and\ \bibinfo {author} {\bibfnamefont {J.}~\bibnamefont
  {Krug}},\ }\href@noop {} {\bibfield  {journal} {\bibinfo  {journal} {Physical
  Review E}\ }\textbf {\bibinfo {volume} {60}},\ \bibinfo {pages} {188}
  (\bibinfo {year} {1999})}\BibitemShut {NoStop}%
\bibitem [{\citenamefont {Schreckenberg}\ \emph {et~al.}(1995)\citenamefont
  {Schreckenberg}, \citenamefont {Schadschneider}, \citenamefont {Nagel},\ and\
  \citenamefont {Ito}}]{schreckenberg1995discrete}%
  \BibitemOpen
  \bibfield  {author} {\bibinfo {author} {\bibfnamefont {M.}~\bibnamefont
  {Schreckenberg}}, \bibinfo {author} {\bibfnamefont {A.}~\bibnamefont
  {Schadschneider}}, \bibinfo {author} {\bibfnamefont {K.}~\bibnamefont
  {Nagel}}, \ and\ \bibinfo {author} {\bibfnamefont {N.}~\bibnamefont {Ito}},\
  }\href@noop {} {\bibfield  {journal} {\bibinfo  {journal} {Physical Review
  E}\ }\textbf {\bibinfo {volume} {51}},\ \bibinfo {pages} {2939} (\bibinfo
  {year} {1995})}\BibitemShut {NoStop}%
\bibitem [{\citenamefont {Barlovic}\ \emph {et~al.}(1998)\citenamefont
  {Barlovic}, \citenamefont {Santen}, \citenamefont {Schadschneider},\ and\
  \citenamefont {Schreckenberg}}]{barlovic1998metastable}%
  \BibitemOpen
  \bibfield  {author} {\bibinfo {author} {\bibfnamefont {R.}~\bibnamefont
  {Barlovic}}, \bibinfo {author} {\bibfnamefont {L.}~\bibnamefont {Santen}},
  \bibinfo {author} {\bibfnamefont {A.}~\bibnamefont {Schadschneider}}, \ and\
  \bibinfo {author} {\bibfnamefont {M.}~\bibnamefont {Schreckenberg}},\
  }\href@noop {} {\bibfield  {journal} {\bibinfo  {journal} {The European
  Physical Journal B-Condensed Matter and Complex Systems}\ }\textbf {\bibinfo
  {volume} {5}},\ \bibinfo {pages} {793} (\bibinfo {year} {1998})}\BibitemShut
  {NoStop}%
\bibitem [{\citenamefont {L{\"u}beck}\ \emph {et~al.}(1998)\citenamefont
  {L{\"u}beck}, \citenamefont {Schreckenberg},\ and\ \citenamefont
  {Usadel}}]{lubeck1998density}%
  \BibitemOpen
  \bibfield  {author} {\bibinfo {author} {\bibfnamefont {S.}~\bibnamefont
  {L{\"u}beck}}, \bibinfo {author} {\bibfnamefont {M.}~\bibnamefont
  {Schreckenberg}}, \ and\ \bibinfo {author} {\bibfnamefont {K.}~\bibnamefont
  {Usadel}},\ }\href@noop {} {\bibfield  {journal} {\bibinfo  {journal}
  {Physical Review E}\ }\textbf {\bibinfo {volume} {57}},\ \bibinfo {pages}
  {1171} (\bibinfo {year} {1998})}\BibitemShut {NoStop}%
\bibitem [{\citenamefont {Krauss}\ \emph {et~al.}(1996)\citenamefont {Krauss},
  \citenamefont {Wagner},\ and\ \citenamefont {Gawron}}]{krauss1996continuous}%
  \BibitemOpen
  \bibfield  {author} {\bibinfo {author} {\bibfnamefont {S.}~\bibnamefont
  {Krauss}}, \bibinfo {author} {\bibfnamefont {P.}~\bibnamefont {Wagner}}, \
  and\ \bibinfo {author} {\bibfnamefont {C.}~\bibnamefont {Gawron}},\
  }\href@noop {} {\bibfield  {journal} {\bibinfo  {journal} {Physical Review
  E}\ }\textbf {\bibinfo {volume} {54}},\ \bibinfo {pages} {3707} (\bibinfo
  {year} {1996})}\BibitemShut {NoStop}%
\bibitem [{\citenamefont {Louhghalam}\ \emph {et~al.}(2013)\citenamefont
  {Louhghalam}, \citenamefont {Akbarian},\ and\ \citenamefont
  {Ulm}}]{deflection}%
  \BibitemOpen
  \bibfield  {author} {\bibinfo {author} {\bibfnamefont {A.}~\bibnamefont
  {Louhghalam}}, \bibinfo {author} {\bibfnamefont {M.}~\bibnamefont
  {Akbarian}}, \ and\ \bibinfo {author} {\bibfnamefont {F.-J.}\ \bibnamefont
  {Ulm}},\ }\href@noop {} {\bibfield  {journal} {\bibinfo  {journal} {Journal
  of Engineering Mechanics}\ }\textbf {\bibinfo {volume} {140}} (\bibinfo
  {year} {2013})}\BibitemShut {NoStop}%
\bibitem [{\citenamefont {Louhghalam}\ \emph {et~al.}(2014)\citenamefont
  {Louhghalam}, \citenamefont {Akbarian},\ and\ \citenamefont
  {Ulm}}]{louhghalam2014scaling}%
  \BibitemOpen
  \bibfield  {author} {\bibinfo {author} {\bibfnamefont {A.}~\bibnamefont
  {Louhghalam}}, \bibinfo {author} {\bibfnamefont {M.}~\bibnamefont
  {Akbarian}}, \ and\ \bibinfo {author} {\bibfnamefont {F.-J.}\ \bibnamefont
  {Ulm}},\ }\href@noop {} {\bibfield  {journal} {\bibinfo  {journal}
  {Transportation Research Record: Journal of the Transportation Research
  Board}\ }\textbf {\bibinfo {volume} {2457}},\ \bibinfo {pages} {95} (\bibinfo
  {year} {2014})}\BibitemShut {NoStop}%
\bibitem [{\citenamefont {Louhghalam}\ \emph {et~al.}(2015)\citenamefont
  {Louhghalam}, \citenamefont {Akbarian},\ and\ \citenamefont
  {Ulm}}]{roughness}%
  \BibitemOpen
  \bibfield  {author} {\bibinfo {author} {\bibfnamefont {A.}~\bibnamefont
  {Louhghalam}}, \bibinfo {author} {\bibfnamefont {M.}~\bibnamefont
  {Akbarian}}, \ and\ \bibinfo {author} {\bibfnamefont {F.-J.}\ \bibnamefont
  {Ulm}},\ }in\ \href@noop {} {\emph {\bibinfo {booktitle} {Transportation
  Research Board 94th Annual Meeting}}}\ (\bibinfo {year} {2015})\BibitemShut
  {NoStop}%
\bibitem [{\citenamefont {Fazeen}\ \emph {et~al.}(2012)\citenamefont {Fazeen},
  \citenamefont {Gozick}, \citenamefont {Dantu}, \citenamefont {Bhukhiya},\
  and\ \citenamefont {Gonz{\'a}lez}}]{fazeen2012safe}%
  \BibitemOpen
  \bibfield  {author} {\bibinfo {author} {\bibfnamefont {M.}~\bibnamefont
  {Fazeen}}, \bibinfo {author} {\bibfnamefont {B.}~\bibnamefont {Gozick}},
  \bibinfo {author} {\bibfnamefont {R.}~\bibnamefont {Dantu}}, \bibinfo
  {author} {\bibfnamefont {M.}~\bibnamefont {Bhukhiya}}, \ and\ \bibinfo
  {author} {\bibfnamefont {M.~C.}\ \bibnamefont {Gonz{\'a}lez}},\ }\href@noop
  {} {\bibfield  {journal} {\bibinfo  {journal} {Intelligent Transportation
  Systems, IEEE Transactions on}\ }\textbf {\bibinfo {volume} {13}},\ \bibinfo
  {pages} {1462} (\bibinfo {year} {2012})}\BibitemShut {NoStop}%
\end{thebibliography}%

\end{document}